# Deducing the key physical properties of a perovskite solar cell from its impedance response: insights from drift-diffusion modelling

Antonio Riquelme[a,#], Laurence J. Bennett[b,#], Nicola E. Courtier[b], Matthew J. Wolf[c], Lidia Contreras-Bernal[a], Alison Walker[c,*], Giles Richardson[b,*], Juan A. Anta[a,*]

[a] Área de Química Física, Universidad Pablo de Olavide, E-41013, Seville, Spain.
[b] Mathematical Sciences, University of Southampton, Southampton, SO17 1BJ, UK.
[c] Department of Physics, University of Bath, Claverton Down, Bath, BA2 7AY, UK

#Both authors contributed equally to this work

**ABSTRACT**

Interpreting the impedance response of perovskite solar cells (PSC) is significantly more challenging than for most other photovoltaics. This is for a variety of reasons, of which the most significant are the mixed ionic-electronic conduction properties of metal halide perovskites and the difficulty in fabricating stable, and reproducible, devices. Experimental studies, conducted on a variety of PSCs, produce a variety of impedance spectra shapes. However, they all possess common features, the most noteworthy of which is that they have at least two signals, at high and low frequency, with different characteristic responses to temperature, illumination and electrical bias. It is shown, by a combination of experiment and drift-diffusion modelling of the ion and charge carrier transport and recombination within the cell, that these common features are well reproduced by the simulation. In addition, we show that the high frequency response contains all the key information relating to the steady-state performance of a PSC, i.e. it is a signature of the recombination mechanisms and provides a measure of charge collection efficiency. Moreover, steady-state performance is significantly affected by the distribution of mobile ionic charge within the perovskite layer. Comparison between the electrical properties of different devices should therefore be made using high frequency impedance measurements performed in the steady-state voltage regime in which the cell is expected to operate.

# Introduction

Perovskite solar cells (PSCs)[1,2] are one of the hottest research topics in contemporary photovoltaics due to the extremely rapid improvements in their performance, which has risen from a record photoconversion efficiency of 9.7% in 2012[3] to one of 25.2% in 2019.[4] Metal halide perovskites are semiconductors with a direct band gap and strong absorption in the visible part of the spectrum.[5] One of their most notable characteristics is that they have a mixed ionic-electronic conduction character, stemming from the high mobility of ionic vacancies in the perovskite structure.[6,7] This chemical property appears to be related to their exceptional optoelectronic properties, in particular, to their excellent photovoltaic charge separation behaviour[8–10], but is also associated with their chemical instability.[11] Further improvements in efficiency and stability necessitate a better understanding of the molecular mechanisms that determine photovoltaic conversion and degradation. In this context, the combination of theoretical modelling, and accurate optoelectronic characterization techniques are a powerful tool.

Impedance spectroscopy (IS) is an electrical characterization technique which analyses the current response of a (photo)electrochemical system to a small periodic voltage modulation. The corresponding impedance response, which is usually assumed to be linear, is measured as a function of the frequency of the modulation and provides information about the kinetic processes occurring within the system. IS has proven to be an indispensable tool for gaining insight into the physical behaviour of solar cells and other photoelectrochemical devices, and has been widely used to characterize organic solar cells and dye-sensitized solar cells.[12] For dye-sensitized solar cells, for instance, a very popular equivalent circuit, devised by Bisquert and co-workers[12–14] has made it possible to extract useful information about the electron dynamics of the device, such as diffusion coefficients, chemical capacitances, electron recombination lifetimes and diffusion lengths. However, the same degree of success has not been achieved for perovskite solar cells. This failure may be primarily attributed to two causes: Firstly, the problem is inherently more complex, involving the motion of both electronic (electrons and holes) and ionic charge carriers, the motions of which are associated with very different time scales, and which interact differently with the internal interfaces within the cell. Secondly, the chemical instability of many of the perovskite materials employed in these cells may lead to a lack of reproducibility of experimental data. In particular, the chemical instability of the perovskite can give rise to exotic features in the impedance spectra of a PSC that are not systematically observed in all configurations and, more frustratingly, may not be observed in all specimens of the same configuration.

The aforementioned complications associated with PSC physics means that, to date, no well-established simple model is to be found in the literature, although some linear equivalent circuits have been proposed. They basically utilise a combination of at least two resistors and two capacitors in various arrangements, -(RC)-(RC)-,[15–18] in order to account for the two main signals that are commonly observed. Some more sophisticated models include additional elements and inductors in an attempt to reproduce exotic features such as loops and negative capacitances.[19–21] However, adding more elements to a given equivalent circuit does not necessarily aid in the interpretation of the measured response of the system, particularly if each equivalent circuit element does not have a well-defined physical meaning. Furthermore, the more elements that are added to a circuit, the more likely it is that an alternative equivalent circuit can be found that provides an identical response; this increases the ambiguity in the physical interpretation of circuit elements, and consequently the utility of the equivalent circuit description is lost.[22] Only very recently a novel equivalent circuit has been proposed that it is inspired by the aforementioned model for dye-sensitized solar cells.[23] However, this equivalent circuit still contains elements attributed to the slow "dielectric" response of the device, which have, as yet, not been established.

Drift-diffusion (DD) models of solar cells, based upon the coupled set of continuity equations for electrons, holes and ions, and Poisson's equation, provides an alternative (and more fundamental) way of understanding the system and its impedance response. As the parameters of DD models such as diffusion coefficients, lifetimes, dielectric constants, doping densities, work functions, etc., have clear physical meanings, a link between the microkinetic processes taking place in the solar cell and its macroscopic characteristics can more readily be established. DD models have successfully been used to simulate the current-voltage (JV) curves of a variety of perovskite solar cells and helped researchers to understand the origin of the hysteresis observed in many cases in terms of dynamic ionic redistribution over the course of the voltage scan.[24–29] Recent developments include models that explicitly account for the influence of the selective contact materials on a cell's behaviour[29] and predictions of the response of the solar cell to small-perturbations in applied voltage and light intensity.[30–32] The latter area is the focus of this paper.

In this context, Jacobs et al.[30] analysed the apparent capacitance of PSCs and its relation to the hysteresis. They used DD modelling to demonstrate that ion-mediated recombination is behind the low frequency "giant" capacitance and the inductive loops observed in the IS data. Moia et al.[29] also used DD modelling to understand the low frequency response in terms of the electronic coupling between ion and electronic motion at the interfaces. Finally, Neukom et al.[30] were able to reproduce, within a single DD framework and with a single parameter set, a wide variety of experimental data for a specific PSC configuration (inverted), including small-perturbation frequency-modulated experiments such as IS at short-circuit. Crucially, none of these works focused on the high frequency impedance response and in particular how it can be used as a tool to deduce the physics associated with the all-important steady-state performance of the cell.

In this paper we use DD numerical simulations in combination with experimental IS to gain a better understanding of the optoelectronic behaviour of PSCs. We show that our DD model can reproduce the main characteristics found in the impedance spectra, and provides a wide-ranging description of the IS features that actually determine the PSC photoconversion efficiency, noting that the main purpose of IS measurements is to detect and quantify loss mechanisms (e.g. slow transport, high recombination rates and limited internal quantum efficiencies) in a particular device. We emphasise that this can only be achieved using a well-defined electrical model, or if particular features of the spectra can be identified with particular loss mechanisms.

The main features of the IS response of PSCs are well summarized in a recent paper by some of the authors.[18] In that work, two devices made from perovskite materials with very different optical properties, were investigated. Nevertheless, it was found that both devices displayed similar impedance responses. The key experimental features of the PSC impedance response are:

(1) The presence of at least two signals or time constants, embodied as arcs in the Nyquist plot and peaks in the frequency plots, one at low frequencies (0.1-10 Hz, LF) and the other at high frequencies ($10^5$-$10^6$, HF).
(2) The LF signal is independent of the DC voltage, whereas the characteristic frequency of the HF response decreases exponentially with increases in the DC voltage.

Furthermore, when an -(RC)-(RC)- equivalent circuit is used to fit the experimental spectrum:

(3) The HF and LF resistances associated to these two signals decrease exponentially with DC voltage, but the so-called "ideality factor" of the device can only be extracted from the slope of the logarithm of the HF (not the LF) resistance versus open-circuit voltage
(4) The HF capacitance is close to being independent of the DC voltage, whereas the LF capacitance increases exponentially with DC voltage, and with the *same* slope as the LF resistance, and hence leads to a voltage-independent LF time constant ($\tau_{LF} = R_{LF} C_{LF}$).

In this work we look for the simplest physical model that can reproduce these four key features of the IS of a PSC. In particular we find that a DD model of the PSC can do so provided that it includes: (1) a three-layered planar configuration including n- and p-type selective contacts and a perovskite (active) layer; (2) electrons in the electron transport layer (ETL); (3) electrons, holes and positively charged ionic vacancies in the perovskite active layer, (4) holes in the hole transport layer (HTL), (5) "rapid" diffusion coefficients for the electron/holes and "slow" diffusion coefficients for the ionic vacancies, respectively, (6) bulk electron-hole recombination in the perovskite layer mediated by a combination of bimolecular and electron-limited Shockley-Read-Hall kinetics and (7) band gaps, work functions, activation energies and dielectric constants in line with values reported in the literature. We then use the "empirical" -(RC)-(RC)- equivalent circuit to fit the numerical IS data and to extract the resistances and capacitances associated with the high- and low-frequency signals. We establish which parts of the spectrum are required to quantify recombination and charge collection and discuss how the ionic subsystem affects the spectrum.

Finally, we show that the key properties of the PSC steady-state performance such as recombination character and charge collection efficiency along with the JV curve are captured by the HF impedance response. In contrast, the LF impedance response is associated with ion motion in the perovskite layer only. The ionic features can, however, affect the HF response and the so-called "ideality factor".

## Numerical and experimental methods

**Numerical method**

JV curves and impedance spectra are numerically simulated using a fully-coupled one-dimensional DD model that accounts for the motion of electrons, holes and positive anion vacancies within a planar perovskite solar cell. The cell is modelled as a three-layer structure in which a perovskite absorber layer is sandwiched between doped electron and hole transport layers. Within the perovskite layer all three charged species are present whilst, within the ETL (HTL), the ion and hole (electron) densities are neglected. Since both holes and electrons are only present in appreciable numbers in the perovskite layer, significant charge carrier recombination only occurs within that e layer, or at its interfaces with the ETL and HTL. Within the perovskite layer these losses are assumed to occur via a combination of the bimolecular and Shockley-Read-Hall (SRH) recombination mechanisms, whilst on the interfaces with the transport layers only SRH recombination is assumed to occur, consistent with the assumption that the free minority carrier density is negligible in the ETL and HTL . Furthermore, significant photogeneration is assumed to occur only within the perovskite and its spatial distribution is modelled by a Beer-Lambert law.

The resulting model of the PSC consists of conservation equations for the three charged species (i.e. electrons, holes and ion vacancies). These couple to the electric potential, within the cell, via Poisson's equation and drift-diffusion laws for the fluxes of the three charged species. Comparison to experiment requires that the model be solved with an appropriate applied potential, $V_{ap}(t)$ (which appears in the form of a boundary condition) that reflects the experimental protocol. We obtain the numerical solution to the model, by using the open-source simulation tool *IonMonger*[33]. This solution can then be used to determine the output current density J(t), on the cell contacts, which, in turn, can be compared to the experimentally determined current flow. This procedure provides a way of interrogating the physical properties of the real cell.

Further details of the model can be found in the works of Richardson *et al*[26] and Courtier *et al*[29], where it was first developed, and a complete description of the modelling assumptions and model equations are given in the Supporting Information. The key model parameters are listed in Table 1.

Impedance spectra are simulated by applying a sinusoidal voltage perturbation of the form

$$V_{ap}(t) = V_0 + V_p \sin(\omega t), \qquad (1)$$

where $V_0$ is the DC voltage, $V_p$ is the amplitude and $\omega$ is the angular frequency of the perturbation. In this procedure the cell is always perturbed about a steady state so that the phase and amplitude of the current response is stable in time. For each frequency, the current response is analysed using a Fourier transform to extract its phase and amplitude. This enables the impedance to be calculated and a spectrum to be constructed by obtaining solutions to the model for 40-100 frequencies over a wide frequency range.

In both the simulation and the experiment (*vide infra*), two situations were considered: (1) for a range of illumination intensities, the voltage was perturbed around the open-circuit voltage consistent with the applied illumination intensity and the parameters are extracted, analysed and plotted as a function of the resulting open-circuit photopotential *V*<sub>oc</sub> and (2) for fixed illumination intensity the voltage was perturbed around a range of applied DC potentials (voltages), and the parameters extracted, analysed and plotted as a function of the DC voltage. In the following, we will use the labels OC and NOC to refer to these two kinds of simulations/experiments.

In the calculations and the experiments we have considered a **regular** configuration in which the ETL and HTL are compact TiO$_2$ and Spiro-OMeTAD, respectively. Results are directly compared to our experimental data (see next subsection). A summary of the parameter set used in the simulations together with the source and justification of each value is collected in Table 1 (general set with no surface recombination) and Table 2 (modified set to fit a particular experimental JV curve). Note that for simplicity the parameters used in Table 1 and 2 are chosen in order to ensure that bulk recombination is electron-limited and the perovskite is effectively p-type. See below for a discussion about the p or n-type character of perovskites and its impact on the modelling.

**Table 1.** Parameter set used in the DD simulations for regular cell configuration with no surface recombination, in which $\epsilon_0$ denotes the permittivity of free space.

| Parameter | Value | Reference / Justification |
|---|---|---|
| Temperature, $T$ | 298 K | |
| Incident photon flux, $F_{ph}$ | 5-1000 W/m$^2$ | It is adjusted to give the desired $V_{oc}$ with the 1-sun equivalent as upper limit |
| **Perovskite Properties** | | |
| Width, $b$ | 3·10$^{-7}$ m | Direct measurement, Ref. [18] |
| Permittivity, $\epsilon_p$ | 24.1 $\epsilon_0$ | Ref. [34] |
| Absorption coefficient at 465 nm, $\alpha$ | 6.5·10$^6$ m$^{-1}$ | Direct measurement (Figure S1) |
| Conduction band minimum, $E_c$ | (-) 3.7 eV | Ref. [35] |
| Valence band maximum, $E_v$ | (-) 5.4 eV | |
| Electron diffusion coefficient, $D_n$ | 5·10$^{-5}$ m$^2$ s$^{-1}$ | Chosen to give a charge collection efficiency close to 100 %, [18,36] together with the electron pseudo-lifetime needed to match the $V_{oc}$ of the experimental JV curve. Very close to the value of 2.05·10$^{-5}$ m$^2$ s$^{-1}$ reported by Wehrenfennig et al.[37] as deduced from experimental value of carrier mobility. |
| Hole diffusion coefficient, $D_p$ | 5·10$^{-5}$ m$^2$ s$^{-1}$ | Similar value to the perovskite electron diffusion coefficient. |
| Conduction band density of states, $g_c$ | 8.1·10$^{24}$ m$^{-3}$ | Ref. [38] |
| Valence band density of states, $g_v$ | 5.8·10$^{24}$ m$^{-3}$ | |
| Mean density of anion vacancies, $N_0$ | 1.6·10$^{25}$ m$^{-3}$ | Ref. [7] |
| Diffusion coefficient for anions, $D_P$ | 10$^{-16}$ m$^2$ s$^{-1}$ | Ref. [6] |
| **Recombination Parameters** | | |
| Electron pseudo-lifetime for SRH, $\tau_n$ | 9·10$^{-7}$ s | Chosen to match the experimental $V_{oc}$. Very close to the value of 7.36·10$^{-7}$ s reported by Zhou et al.[25] |
| Hole pseudo-lifetime for SRH, $\tau_p$ | 3·10$^{-9}$ s | Two orders of magnitude smaller than the electron pseudo-lifetime, therefore this parameter does not affect the calculations when the perovskite is p-type (i.e. when there is a greater density of holes than electrons in the perovskite layer). |
| Bimolecular recombination parameter, $\beta$ | $9.4 \cdot 10^{-16}$ m³ s⁻¹ | Ref[37] |
| **ETL properties** | | |
| Fermi level (work function), $E_{fE}$ | (-) 4.0 eV | Ref.[24] |
| Effective doping density, $d_E$ | $1.1 \cdot 10^{26}$ m⁻³ | Ref.[39] |
| Width, $b_E$ | $1 \cdot 10^{-7}$ m | Direct measurement |
| Permittivity, $\epsilon_E$ | $20\,\epsilon_0$ | Ref.[40] |
| Electron diffusion coefficient, $D_E$ | $5.14 \cdot 10^{-7}$ m² s⁻¹ | Ref.[24] |
| **HTL Properties** | | |
| Fermi level (work function), $E_{fH}$ | (-) 5.22 eV | Ref.[24] Chosen to give a small energy offset to the perovskite valence band (0.18 eV) to ensure that there exists a relatively large density of holes, compared to the electron density, within the perovskite, ensuring electron-limited recombination |
| Effective doping density, $d_H$ | $1.1 \cdot 10^{26}$ m⁻³ | Similar value to the effective doping density of ETL. |
| Width, $b_H$ | $3 \cdot 10^{-7}$ m | Direct measurement. Ref.[18] |
| Permittivity, $\epsilon_H$ | $3\,\epsilon_0$ | Ref.[24] |
| Hole diffusion coefficient, $D_H$ | $2.57 \cdot 10^{-9}$ m² s⁻¹ | Ref.[41] |

**Table 2.** Parameters changed in the DD simulations in order to fit the experimental data of this work (only parameters which are different from Table 1 are indicated).

| Parameter | Value | Reference / Justification |
|---|---|---|
| **Perovskite Properties** | | |
| Conduction band density of states, $g_c$ | $8.1 \cdot 10^{25}$ m⁻³ | Modification introduced to match the $V_{oc}$ without affecting the fill factor.[38] |
| Valence band density of states, $g_v$ | $5.8 \cdot 10^{25}$ m⁻³ | |
| Mean density of anion vacancies, $N_0$ | $1.6 \cdot 10^{26}$ m⁻³ | Modification introduced to reproduce the hysteresis of the experimental JV curve.[7] |
| Electron diffusion coefficient, $D_n$ | $8 \cdot 10^{-6}$ m² s⁻¹ | Modification introduced to match the experimental CCE after changing the electron pseudo-lifetime |
| Hole diffusion coefficient, $D_p$ | $8 \cdot 10^{-6}$ m² s⁻¹ | |
| Diffusion coefficient for anions, $D_P$ | $3 \cdot 10^{-17}$ m² s⁻¹ | Modification introduced to reproduce the hysteresis at the same scan-rate of the experimental JV curve.[6] |
| **Recombination Parameters** | | |
| Electron pseudo-lifetime for SRH, $\tau_n$ | $2 \cdot 10^{-7}$ s | Chosen to match the experimental $V_{oc}$. Very close to the value of $7.36 \cdot 10^{-7}$ s reported by Zhou et al.[25] |
| Electron recombination velocity at ETL, $vnE$ | $1 \cdot 10^5$ m s⁻¹ | Introduced to reproduce the Ideality Factor and the Fill Factor of the experimental JV curve. |
| Hole recombination velocity at ETL, $vpE$ | 20 m s⁻¹ | Introduced to reproduce the Ideality Factor and the Fill Factor of the experimental JV curve. |
| Electron recombination velocity at HTL, $vnH$ | 7.5 m s⁻¹ | Introduced to reproduce the Ideality Factor and the Fill Factor of the experimental JV curve. |

| Hole recombination velocity at HTL, $v_{pH}$ | $1\cdot 10^5$ m s$^{-1}$ | Introduced to reproduce the Ideality Factor and the Fill Factor of the experimental JV curve. |

Simulated JV curves and impedance spectra are obtained using the same measurement protocol as used in the experiment, described below. To be consistent with the experiments (see below), all JV curves were generated with a scan rate of 100 mV/s after 20 seconds of pre-conditioning at 1.2 V. Impedance data were extracted for a 20 mV perturbation in the 0.01-10$^6$ Hz range. The illumination considered in both the JV curve generation and the IS simulation was blue monochromatic light ($\lambda$ = 465 nm) fixed at varying intensities in the 5-1000 W/m$^2$ range.

**Experimental methods**

Numerical simulations are compared to experimental results taken from the literature[15–19,42,43] and to devices, with a regular configuration (TiO$_2$/MAPbI$_3$/Spiro-OMeTAD), that were specifically fabricated in our lab. See Supporting Information for fabrication details.

Experimental current-voltage characteristics of the devices were obtained using a solar simulator (ABET-Sun2000) under 1000 W/m$^2$ illumination with an AM 1.5G filter. The light intensity was recorded using a reference mono-crystalline silicon solar cell with temperature output (ORIEL, 91150). A metal mask was used to define an active area of 0.16 cm$^2$. The current-voltage characteristics were then determined by applying an external potential bias to the cell and measuring the photocurrent using an Autolab/PGSTAT302N potentiostat. The current-voltage characteristics were measured with a scan rate of 100 mV/s and a sweep delay of 20s.

The illumination for the IS measurements was provided by blue LED ($\lambda$ = 465 nm) over a wide range of DC light intensities (10-500 W/m$^2$). As indicated above, two types of IS experiments were analysed: (1) at open-circuit (OC) and (2) at non open-circuit (NOC) conditions.[18] In the latter case the parameters are corrected for voltage drop arising from the induced DC current and the corresponding series resistance. In both OC and NOC conditions a 20 mV (peak) perturbation in the 10$^{-2}$-10$^6$ Hz range was applied. A response analyser module (PGSTAT302N/FRA2, Autolab) was utilized to analyse the frequency response of the devices.

# Results and discussion

## JV characteristics and impedance spectra shapes

Simulation results for a perovskite solar cell with regular configuration, under blue light illumination, obtained using the DD model discussed above and the parameter set given in Table 1 (no surface recombination), and Table 2 (with surface recombination) are presented in **Figure 1**. In Figure S1 in the Supporting Information we provide the experimental photovoltaic performance and optical absorption data under simulated AM1.5G 1000 W/m² illumination of the studied device. The experimental JV curve yields a $V_{oc}$ > 1 V and 15% average power conversion efficiency. This performance is close to state-of-the-art values for $TiO_2$/$MAPbI_3$/Spiro-OMeTAD devices.[18,44]

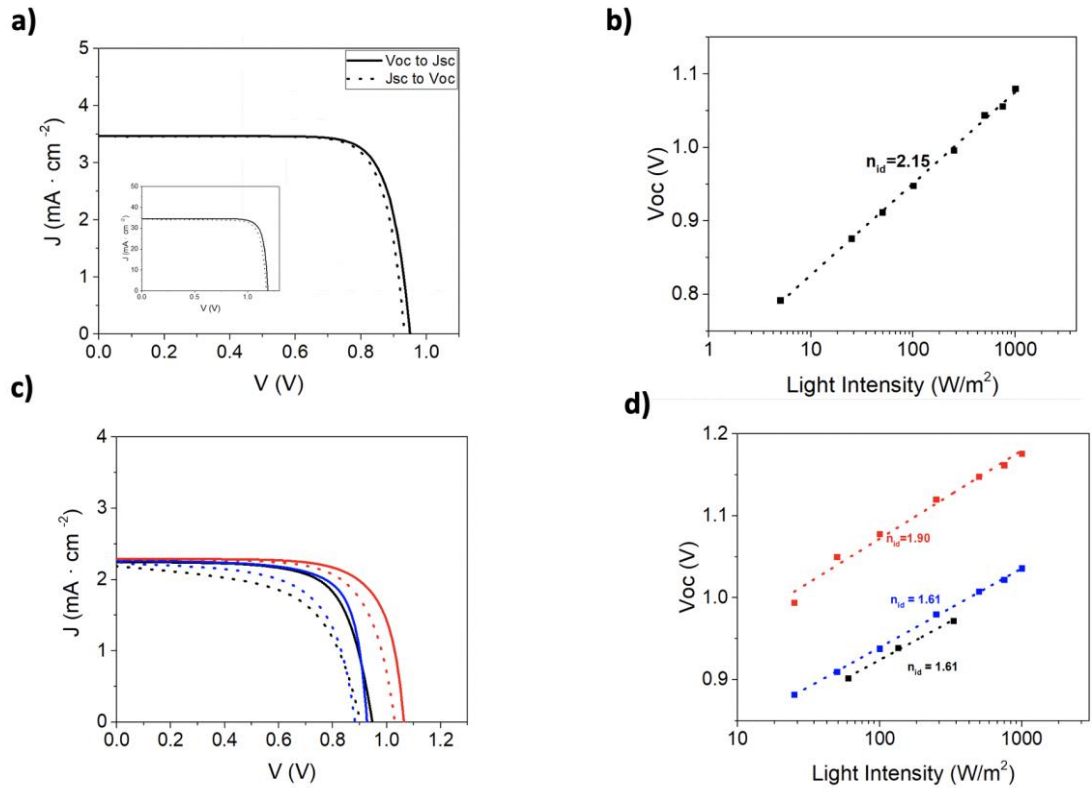

**Figure 1.** Simulated steady-state results under blue light illumination: JV curve **(a)** and open-circuit voltage vs. light intensity **(b)** for the parameters indicated in Table 1. The JV curve in (a) is displayed at two light intensities: 100 W/m² and 1000 W/m² (inset). Simulation (red and blue lines) and experimental (black line) results of a $TiO_2$/$MAPbI_3$/Spiro-OMeTAD solar cell with ideality factor different from 2: **(c)** current-voltage curve with 100 mV/s scan rate under blue light illumination and **(d)** semilogarithmic plot of photovoltage versus light intensity. The parameters used in the simulation in (c) and (d) are those of Table 2. Dashed lines in (b) and (d) are linear fits to Eq. (8).

Numerical results in Figure 1 confirm that the numerical DD model is capable of reproducing the JV characteristics as well as the $V_{oc}$-ln($I$) plot with an "ideality factor" that is close to 2, as observed in many experimental reports for cells with a regular configuration.[17,18,43,45,46] However, although incorporation of surface recombination into the model reduces the value of the ideality factor, it does not accurately reflect the recombination mechanism as it would, for example, for an np-junction; this is a consequence of the presence of mobile ions in the perovskite. Indeed, it can be shown that changing the ionic density, while keeping the recombination mechanism unaltered, results in a significant change to the ideality factor (*vide infra*).

In **Figure 2** we present simulated impedance spectra for a regular cell configuration (with parameter set given in Table 1) at open-circuit conditions. In Figure S2 of the Supporting Information we show additional impedance spectra at lower light intensities for the same parameter set. Two signals in the impedance spectra are obtained in the form of arcs in the Nyquist plot (in which the imaginary part of the impedance $Z''$ is plotted versus the real part $Z'$) and in the form of peaks in the Cole-Cole plot ($Z''$ is plotted versus frequency). The high frequency peaks appear at frequencies of around $10^5$-$10^6$ Hz, whereas the low frequency peaks occur below 1 Hz. The shape of the spectra, the way it changes shape with modification of the illumination intensity and the positions in the frequency domain of the maximum response are all consistent with what is typically seen in experimental impedance measurements of perovskite solar cells, in both regular and inverted configurations.[15–19,22,42,43] For instance, the size of the HF arc decreases and its frequency increases with illumination intensity, whereas the maximum of the LF signal remains almost unchanged in its position in the x-axis (Figure 2b).

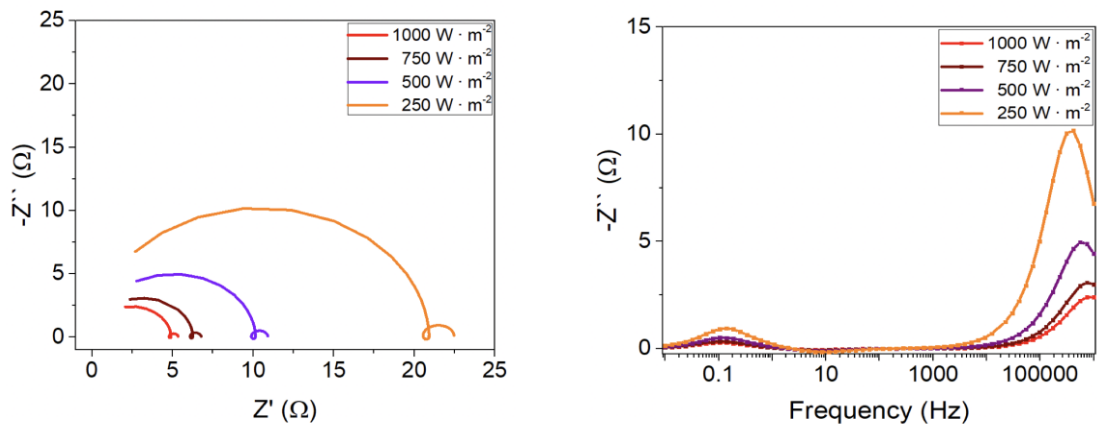

**Figure 2.** Simulated impedance spectra at open-circuit conditions and variable illumination intensity for the parameters indicated in Table 1. **Left:** Nyquist plot. **Right:** Cole-Cole plot.

**HF and LF resistances and "ideality factor"**

Numerical values associated with the high and low frequency component of the resistance and the capacitance need to be extracted by fitting to an equivalent circuit. Knowing that there are two well defined frequency signals, the simplest (empirical) description is based on a Voight model with two RC elements in series (-(RC)-(RC)-). Alternatively, a "nested" arrangement can be used, which actually provides the same results[22] (Figure S3). Note that this equivalent circuit does not have any particular physical meaning. It is just an instrumental procedure to extract resistance and capacitance values from the two time signals observed in both the simulated and the experimental spectra.

Results from the simulations performed at open circuit (at $V=V_{OC}$), and with no series resistance ($R_{ser}$), are shown in **Figure 3**. In line with widely reported experimental trends[16–19,42,43], both high and low frequency resistances exhibit an exponential dependence on the open-circuit voltage $V_{oc}$, (which is a function of the incident light intensity). However, in contrast to some reports[16,43], the slopes of $\ln(R_{HF})$ and $\ln(R_{LF})$, although similar, are not equal. In fact, as explained below (Eq. 7), only the slope of $\ln(R_{HF})$ (the logarithm of the high frequency resistance) when plotted against $V_{OC}/k_BT$ should should be interpreted as $n_{id}$, the "ideality factor" of the solar cell. Hence, we claim that $\ln(R_{LF})$ cannot be used to extract $n_{id}$ as inferred from previous work.[16,43] $n_{id}$ is usually defined to be the slope of $V_{OC}/k_BT$ plotted against the log of the light intensity (i.e. $\ln(I)$), that is

$$I = I_o \exp\left(\frac{V_{OC}}{n_{id}k_BT}\right) \qquad (2)$$

(see Figure 1b, and Eq. 8). Notably we find that in this, and in other simulations that we have conducted, that these two different ways of computing $n_{id}$ yield the same result. Indeed it has been found that that both methods of interpreting experimental data yield the same ideality factor, as, for example, reported on for TiO$_2$(c)/TiO$_2$(mp)/MAPbI$_3$/Spiro-OMeTAD devices.[17,18]

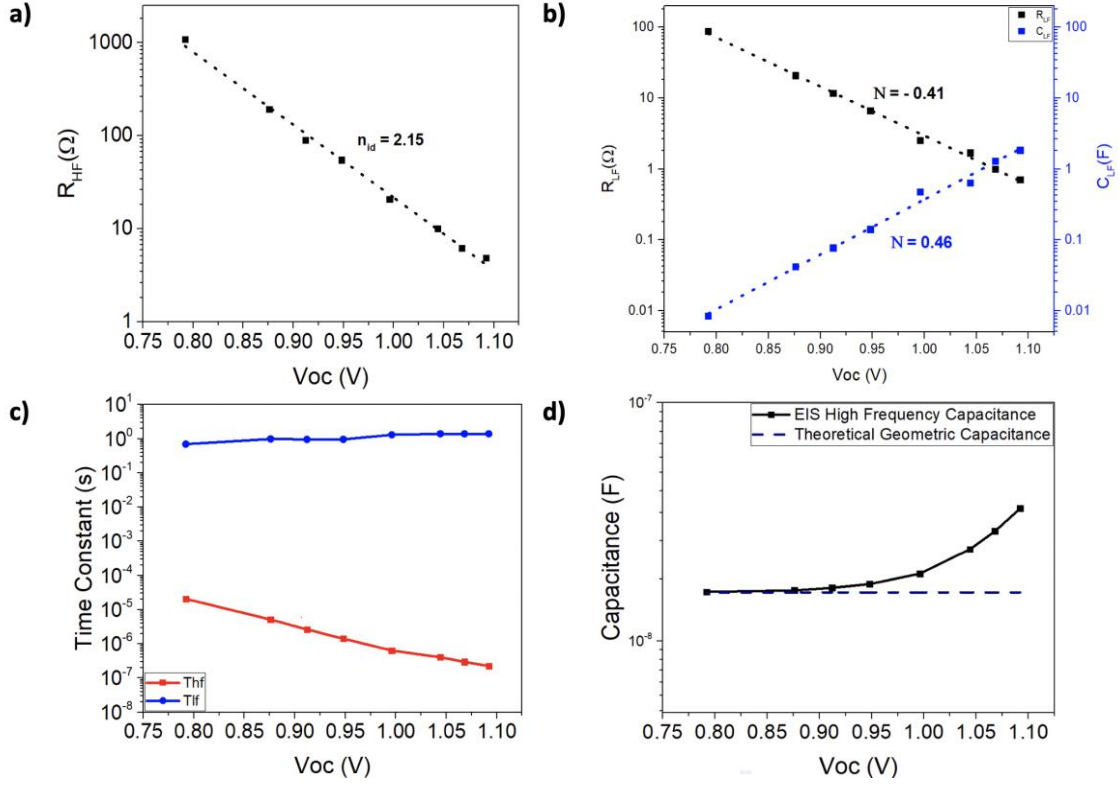

**Figure 3.** Equivalent circuit parameters extracted from DD simulation of the impedance response for the parameters indicated in Table 1. **(a)** High frequency resistance, with the dashed line a linear fit to Eq. (7). **(b)** Low frequency resistance and capacitance, with the dashed lines standing for linear fits to Eqs. (10) and (11). **(c)** High and low frequency time constants as obtained from the frequency peaks in Cole-Cole plot **(d)** High frequency capacitance and theoretical value of the capacitance as predicted from Eq. (9).

In order to provide a simple explanation as to why $\ln(R_{HF})=n_{id} V_{OC}/k_B T$ +const., where $n_{id}$ is as defined in Eq.(2), we make the following simple argument. In line with Contreras-Bernal et al.[18] we express the bulk recombination current in terms of the bulk electron density $n$ using the kinetic expression

$$J_{rec} \approx -qb\frac{dn}{dt} = qbk_{rec}n^\gamma \qquad (3)$$

where $k_{rec}$ is the recombination kinetic constant, $\gamma$ is the reaction order, and $b$ is the perovskite film thickness.

Eq. (3) describes electron-limited recombination. Whether perovskites are n-type or p-type is a long debate and depends strongly on the substrate, the particular perovskite formulation and the preparation conditions.[40,47] As noted above, for the particular case studied here, as described by the parameters in Table 1, the disparity between the electron and hole quasi-lifetimes (see Table 1, $\tau_n = 100\tau_p$) and the disparity between the energy band offset at each interface ($E_c - E_{fE}$ = 0.3 eV while $E_{fH} - E_v$ = 0.18 eV), means that the electron density in the perovskite layer is relatively small and long-lived, compared to the hole density. Consequently, the recombination rate is limited by the electron density. However, an analogous expression to Eq. (3), in terms of hole density, would hold for hole-limited recombination, as would the simplified description below.

On making the assumption that the electrons and holes are in quasi-equilibrium, the usual assumption when deriving a diode equation, we obtain the relation[17,48]

$$np = N_c N_p exp\left[-\frac{(E_c-E_v)-(E_{F,n}-E_{F,p})}{k_B T}\right] = N_c N_p exp\left[-\frac{E_g-qV}{k_B T}\right] \quad (4)$$

where $E_g$ is the band gap, $V$ is the applied voltage and $N_c$ and $N_v$ are the effective densities of conduction band and valence band states. Since the electrons and holes are not the only charged species, indeed they typically occur at considerably lower densities that the ion vacancies, we cannot appeal to charge neutrality to assert $p \approx n$. However the density of these two species are usually correlated and so it is not unreasonable to expect that they grow via a power relation $p \approx n^{\alpha-1}$, for some constant α, such that

$$n = (N_c N_p)^{1/\alpha} exp\left[-\frac{E_g-qV}{\alpha k_B T}\right] \quad (5)$$

Thus, under these conditions, the recombination current can be related to the applied voltage, using (3) and (5), via the equation

$$J_{rec} = qdU_{rec} = J_{00} exp\left[-\left(\frac{\gamma}{\alpha}\right)\frac{E_g-qV}{k_B T}\right] \quad (6)$$

The derivative of this equation with respect to voltage gives the recombination resistance:

$$R_{rec} = \left(\frac{\partial J_{rec}}{\partial V}\right)^{-1} = R_{00} exp\left[-\frac{qV}{n_{id}k_B T}\right]; \quad \text{with} \quad n_{id} = \alpha/\gamma \quad (7)$$

On imposing open circuit conditions $V = V_{OC}$, for which the total charge generation rate across the cell $G$ is equal to the recombination rate, we find that

$$V_{OC} = \frac{E_g}{q} - \frac{n_{id}k_B T}{q} \ln\left(\frac{J_{00}}{qdG}\right) \quad (8)$$

where $n_{id}$, the so-called ideality factor, is given, as indicated, by $n_{id} = \alpha/\gamma$. Assuming a reaction order of $\gamma \sim 1$ (a limiting case of SRH recombination), and $\alpha \sim 2$, we would predict an ideality factor of $n_{id} \sim 2$. Eq. (8) also predicts a linear dependence of $V_{OC}$ versus absolute temperature, with $V_{OC} \to E_g/q$, when $T \to 0$. This is another common experimental result[17,45,49] and is also approximately reproduced by the DD numerical simulation (Figure S4).

It is important to stress that the "ideality factor" is dependent, not only on the recombination mechanisms, but also on the ion vacancy concentration. This is illustrated in the Supporting Information where we re-calculate $n_{id}$ for different ion vacancy concentrations (in Figure S5), finding different values despite the recombination mechanism being unchanged. The discrepancy between the band gap predicted by the extrapolation of the open-circuit voltages from our DD simulations (1.60 eV) and the band gap defined by the parameter set in Table 1 (1.7 eV) is, however, a result of our ad-hoc assumptions about the charge-carrier distributions. In a forthcoming work we aim to provide a more sophisticated approximate analytic model with which to interpret data.

Still, the simple model described by Eqs. (6)-(8) demonstrates that the slope of the recombination resistance and the $V_{oc}$-ln($I$) plot are connected, a fact that is borne out both in the simulations (Figures 3a, 1b, and S5) and experiment.[17,18] with the proviso that the high frequency resistance is identified as the true recombination resistance. ($R_{rec}$ = $R_{HF}$). A further confirmation of this result will be discussed below. The fact that the DD model, described in the Methods section, reproduces the experimental observations is a notable result and implies that the electronic system can be decoupled from the ionic system at high frequency (HF). This allows recombination parameters to be extracted directly from the HF IS data, as it is the case for most of other types of solar cell. However, evidence that both electronic and ionic signals are still connected is shown in **Figure 4** and explained below.

**HF and LF capacitances, and time constants**

In contrast to low and high frequency resistances, the low and high frequency capacitances exhibit markedly different behaviours. The high frequency capacitance (Figure 3d) is voltage-independent until $V_{OC}$ exceeds around 1V. In fact, the simulated value in the "flat" zone coincides very well with the geometrical capacitance of the device

$$C_g = \frac{A\varepsilon_r\varepsilon_0}{b} \approx 1.6\ 10^{-8}\ F \qquad (9)$$

where $A$ is the geometrical area and $b$ is the perovskite film thickness as specified in Table 1. Flat capacitances are also observed in the experiments, even for perovskite solar cells with mesoporous $TiO_2$ contacts. In the experimental case a roughness factor of around 100 has to be introduced into Eq. (9) to match the value yielded by the IS analysis.[50] The increase of the capacitance at high voltages is due to accumulation of charges in the bulk. This behaviour has been recently discussed by Kiermasch et al.[51]

The theoretical results from the DD model (Figure 3b) also show the low frequency capacitance to be an exponentially increasing function of $V_{OC}$ which is in agreement with experimental findings.[18,43] Notably the slope of the ln($C_{LF}$) vs $V_{OC}$ curve is almost exactly the reciprocal of the slope of the ln($R_{LF}$) vs $V_{OC}$ curve

$$R_{LF} \sim exp\left[-N\frac{qV}{k_BT}\right] \quad (10)$$

$$C_{LF} \sim exp\left[N\frac{qV}{k_BT}\right] \quad (11)$$

with N ~ 0.41-0.46. Consequently, we are able to define a low frequency time constant

$$\omega_{LF} = 1/(R_{LF}C_{LF}) \quad (12)$$

which remains constant as the illumination, and hence $V_{OC}$, is adjusted. This feature, which is systematically observed in the experiments[18,43] is here exactly reproduced by the DD model (Figures 3b and 3c). Again, this demonstrates that the slow ionic motion of a single positive species coupled to the electronic motion of electrons and holes is the only microscopic assumption that is required to reproduce the low frequency behaviour. The ionic resistance and capacitance have been described by Moia et al.[31] in terms of transport ionic resistances and interfacial capacitances. However, this is not an unambiguous demonstration that the slopes of their natural logarithms with respect to voltage should be the same. Furthermore, to the best of our knowledge, no other explanation has been presented in the literature. In principle, an ionic resistance should be a function of the ionic diffusion coefficient only, whereas the interfacial capacitance depends upon the accumulation of charges at the interface. An accumulation model that explains the exponential dependence of the low frequency capacitance on $V_{OC}$ has been proposed.[52] However, the coupling observed between the low frequency resistance and capacitance has remained unexplained. An "ionically modulated recombination" has also been proposed[16,29,30], but we have demonstrated above that only the high frequency resistance yields the correct value of the "ideality factor", suggesting that the recombination mechanisms can only be deduced from the high frequency signal. This is because the slow-moving ion vacancies are unaffected by the high frequency signal and so remain in the steady-state configuration that they adopt when the cell is used to generate current. This mechanism is discussed in detail below.

**Impact of ionic distributions on the LF and HF signals**

To shed more light on the nature of the low frequency signal, we have simulated the impedance spectra at open circuit, using the DD model, for various sets of the ionic parameters but keeping everything else fixed. The results of these simulations are shown in **Figure 4**. The results clearly demonstrate that modification of the ionic diffusion coefficient (while keeping the density of vacancies fixed) shifts the frequency at which the low frequency peak is observed, whereas the high frequency peak remains unaltered. Modification of the density of vacancies (while keeping the diffusion coefficient fixed) does not shift the low frequency peak but appreciably reduces the intensity of the HF signal and increases that of the LF signal.

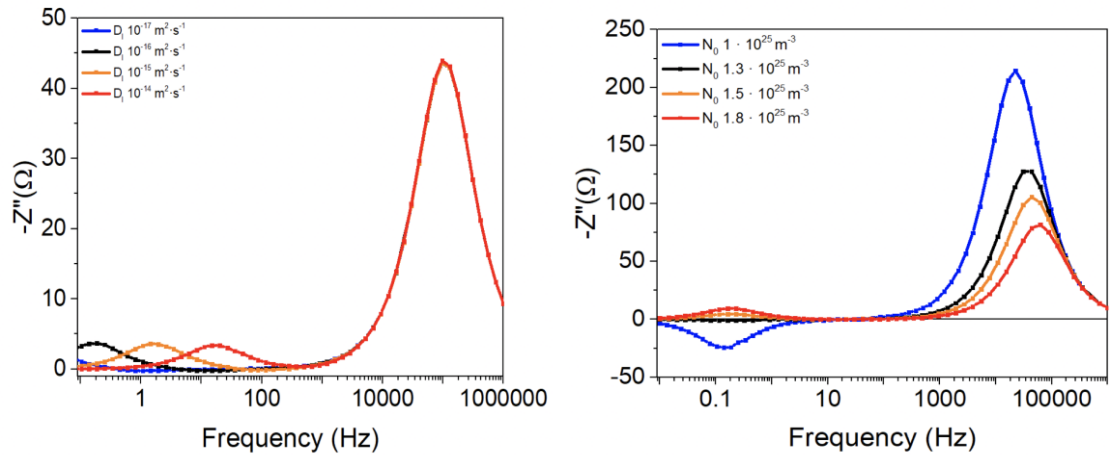

**Figure 4.** Frequency plots at open circuit for varying values of the ionic vacancy diffusion coefficient and the density of vacancies.

Vacancy concentrations above $1.3 \cdot 10^{25}$ m$^{-3}$ produce two peaks at high and low frequencies whereas values below this threshold decreases the height of the low frequency signal and, for sufficiently low defect concentration, can lead to a negative peak height. Indeed, this is the negative signal that is responsible for the apparent inductive loops that can appear in the Nyquist plots of the IS. Concentrations of around $2 \cdot 10^{25}$ m$^{-3}$ for charge-neutral iodine vacancies have been predicted for MAPbI$_3$ by Walsh et al.[7] Defect formation energies calculated by Yang et al.[53] also yield values around $10^{26}$ m$^{-3}$. In contrast, Bertoluzzi et al.[54] obtained $4 \cdot 10^{23}$ m$^{-3}$ for MAPbI$_3$ using an analytical DD model. The wide range of possible ion vacancy densities can be used to explain why, for some materials such as CsPbBr$_3$ which have different defect formation energetics to MAPbI$_3$, the low frequency signal may not appear[18] and why certain specimens show negative signals and others do not, since the ionic defect density depends strongly on temperature[55] and the experimental preparation of the device.

The trend of decreasing LF peak (and increasing HF peak) with decreases in the ion vacancy density can be explained if we consider what fraction of the total potential drop occurs across the perovskite layer (as opposed to the electron and hole transport layers). It must be born in mind that the impedance signal is the response of the collection of electronic carriers to an electric potential perturbation , which depends not only on the applied voltage but also on the distribution of ions in the device. For relatively high ion vacancy concentration in the perovskite, the potential drop occurs mainly across the transport layers which act to screen most of the electric field from the perovskite layer (as explained in Courtier et al.[29]). In contrast, for low ion vacancy concentrations, most of the potential drop occurs in the perovskite layer, and this gives rise to large internal electric fields (within the perovskite layer) that vary significantly with changes in the applied voltage. In the case where the predominant loss mechanism is recombination within the perovskite material, these electric fields determine the collection efficiency, by either driving carriers towards, or away from, the appropriate contacts.

In Figures S6 and S7 the electric potential and ionic vacancies density profiles at steady state are plotted at different DC voltages for the standard parameter set of Table 1. As previously demonstrated[26,29] ionic charge accumulates in narrow layers, called Debye layers (or space-charge layers), adjacent to the perovskite interfaces until, at steady state, the electric field in the bulk of the perovskite layer is completely screened. At steady state, all of the potential difference across a PSC occurs across the Debye layers. Additionally, due to the high density of mobile ionic charges, the electric potential across the perovskite layer of a PSC is determined almost solely by the distribution of ions (and not the relatively small density of electronic charges).[26,29] The resulting electric potential distribution controls the distribution of electronic charge carriers across the PSC and the potential barriers which aid or hinder their extraction from the device.

In **Figure 5** the variation of the ionic distribution and the electric potential during the impedance experiment is shown. It can be observed that at high frequencies the voltage perturbation occurs mainly in the perovskite layer whereas at low frequencies the electric potential in the perovskite remains unaltered. This is easily understood bearing in mind that at high frequencies ions do not have time to move and therefore remain fixed in their steady-state distribution at that applied DC voltage. Consequently, the potential drops across these four Debye layers also remain fixed. It follows that the periodic perturbations in the applied voltage, induced by the IS protocol, leads to a corresponding periodic change in voltage drop across the central bulk region of the perovskite layer; this manifests itself as a uniform time-periodic electric field that permeates this bulk region. In summary, the HF voltage perturbation leads to an alternating positive/negative electric field (negative/positive slope in Figure 5 left) across the bulk of the perovskite layer that causes enhanced/reduced extraction of the electronic carriers. This fluctuating recombination current is hence in-phase with the voltage perturbations. The out-of-phase component of the impedance response comes from capacitive contributions either at the interfaces (geometrical capacitance)[19,42], Eq. (9), or in the bulk (chemical capacitance)[51], high voltage region in Figure 3d. Consequently, $Z'' < 0$. This is the origin of the HF peak in Figures 2, 4, and S2.

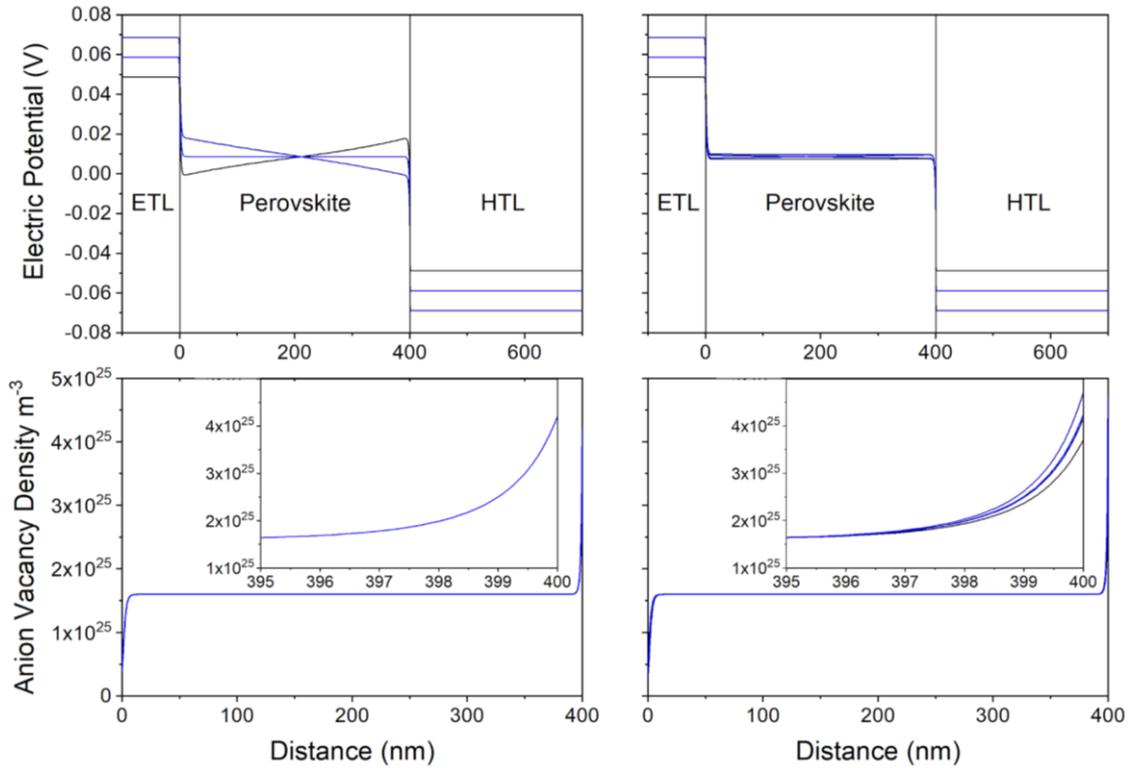

**Figure 5. Top.** Evolution of the anion vacancy distribution over a frequency period under illumination (1000 W/m$^2$) at V = 1.1 V with the standard parameter set of Table 1. Results at both high (10$^4$ Hz, left) and low (10$^{-2}$ Hz, right) frequencies are shown. Insets show the right Debye layer within the perovskite at the perovskite/HTL interface. **Bottom.** Evolution of the electric potential over a frequency period in the impedance experiment under illumination (1000 W/m$^2$) at V = 1.1 V with the standard parameter set of Table 1. Results at both high (10$^4$ Hz, left) and low (10$^{-2}$ Hz, right) frequencies are shown.

In contrast, at low frequencies the ions *do* have sufficient time, over the period of the voltage perturbation, to migrate in and out of the Debye layers at the interfaces. The potential drops occurring across these Debye layers can thus adjust quickly enough, over the course of a single period, to offset some of the perturbation in the applied voltage across the cell. Once again, the perturbations in the applied voltage cause increases/decreases in the electric field across the bulk of the perovskite but their size is reduced because some of the perturbation in the applied voltage now occurs across the Debye layers. Indeed, at sufficiently low frequencies, the Debye layers have sufficient time to fully adjust to the perturbation in the applied voltage and so the electric field in the bulk of the perovskite is zero. Reorganization of ions in the Debye layers at LF voltage perturbations cause increases/decreases in the electric field across the bulk of the perovskite which correspond to decreases/increases in the sizes of the potential drops across the Debye layers (i.e. positive bulk fields correspond to under-filled Debye layers which are charging, while negative bulk fields correspond to overfilled Debye layers which are discharging). Therefore, the LF voltage perturbations have opposing effects on recombination, which changes from being in-phase at HF to out-of-phase at LF. Hence, at low frequencies, bulk recombination in combination with large internal electric fields can cause a negative response, while the effects of the interfaces cause a positive response. This is why, in Figure 4b, the smallest vacancy density (with the largest internal electric fields) produces both the highest HF peak and the negative LF peak.[29] Hence, we demonstrate that it is possible to see negative impedance from a PSC limited by bulk recombination, when the density of mobile ionic charge in its perovskite layer is sufficiently low.

Importantly, as both LF resistance and capacitance contributions arise from the charging/discharging of the same interfacial ionic distribution the voltage dependence is the same. This explains the opposite slopes for the LF components of recombination and capacitance in Figure 3b. It also explains the shift of the peak towards higher frequencies when the ions have a higher diffusion coefficient (which allows faster charging/discharging of the Debye layers). In contrast to the interpretation by Moia et al.[31] the low frequency resistance is not directly an ionic transport resistance but a recombination resistance modulated by ion transport from the bulk to/from the Debye layers.

**HF signal and charge collection efficiency at steady-state**

In the discussion above we state that only the HF signal provides information about recombination and the "ideality factor". To confirm this, we have run impedance simulations with fixed illumination and a variable external DC voltage (NOC conditions), in order to recreate the extraction of charge and the generation of a photocurrent along the JV curve. The resistances and capacitances obtained from the fittings to the double RC equivalent circuit (Figure S3) as a function of the external DC voltage are shown in **Figure 6.**

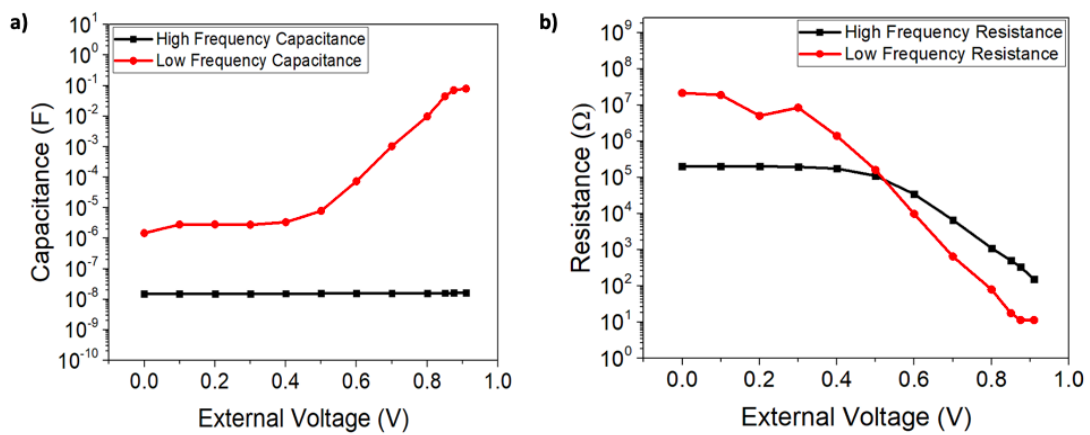

**Figure 6.** Capacitance and resistance values extracted from the equivalent circuit of Figure S3 and the simulated impedance spectra with parameters from Table 1 at non open-circuit conditions: constant 100 W/m² illumination and variable external DC voltage.

The most visible difference with respect to OC conditions is that the low and high frequency resistances, as well as the low frequency capacitance, only vary exponentially with applied voltage in the vicinity of the maximum power point and close to open circuit. At conditions close to short-circuit, these magnitudes reach a plateau and remain effectively constant down to $V$ = 0. In contrast, the high frequency capacitance retains its geometrical character at all voltages. This behaviour is completely consistent with what is observed in IS experiments of perovskite solar cells[18,36,43], further confirming that the model utilized with the set of parameters in Table 1 is enough to reproduce the overall impedance response of the cell.

The following expression has been proposed to estimate the charge collection efficiency from impedance data[18,43]

$$CCE \approx 1 - \frac{R_{rec(OC)}}{R_{rec(NOC)}} \qquad (13)$$

In **Figure 7** we plot the collection efficiency predicted by this equation when we assume that the high frequency resistance is the recombination resistance, $R_{rec} = R_{HF}$. We base this assumption on the recombination character of the high frequency signal, as indicated by the discussions above.

As expected, CCE is zero at the open-circuit photovoltage (see Figure 1) and reaches a maximum at short circuit. In fact, a value of 100% is predicted at short-circuit conditions, in line with measured values for state-of-the-art devices.[18,43,46,56] The result of Eq. (13) can be compared with a direct "measurement" of the internal quantum efficiency (IQE), defined as the ratio between the photocurrent density and the absorbed photon current density (absorbed photon flux per unit charge). Knowing that in the DD simulation each absorbed photon produces one and only one electron-hole pair, these two efficiencies should be identical. **Figure 7** demonstrates that this is indeed the case. Furthermore, the agreement between the two ways of calculating the CCE/IQE confirms that the high frequency resistance is a *true* recombination resistance, in line with the coincidence between its voltage slope and the "ideality factor".

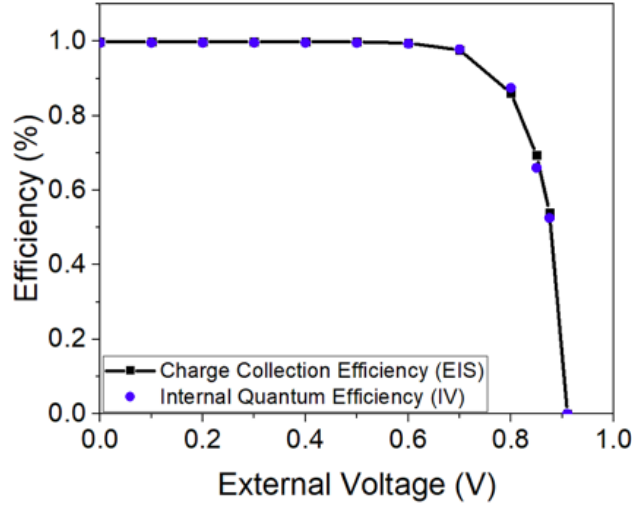

**Figure 7.** Simulated charge collection efficiency as predicted from Eq. (13) and the impedance parameters of Figure 6b when we assume the identity $R_{rec} = R_{HF}$ (black line). The stationary internal quantum efficiency derived from the simulated JV curve (current density / $q$ · absorbed photon flux) is included in the graph (blue circles). Simulation parameters are those in Table 1.

In a recent experimental work, we carried out a similar analysis of the collection efficiency for MAPbI$_3$ based devices.[18] In that case the agreement between the result of Eq. (13) (with $R_{rec} = R_{HF}$) and the IQE derived from the experimental photocurrent was only approximate. Although the coincidence between the two magnitudes is improved for devices of better efficiency, there is always a difference between the two, indicating that in real devices the connection between both magnitudes is only approximate, $R_{rec} \sim R_{HF}$. In Figures S8 and S9 in the Supporting Information simulated results for the impedance and the charge collection efficiency with data taken from Table 2 are presented. In this case, as mentioned above, a certain amount of surface recombination is needed to match the experimental "ideality factor" and to get an improved reproduction of the experimental impedance spectrum. The result is that in this case the coincidence between the prediction of Eq. (13) and the IQE is no longer exact.

## Conclusions

In this work we have used a drift-diffusion model, which explicitly accounts for the motion of electrons, holes and positive ion vacancies, to interpret impedance spectroscopy data of perovskite solar cells with an ETL/Perovskite/HTL architecture. The main result of the work is to provide a recipe with which to interpret the impedance response data in terms of the physics of its steady-state operation. From a practical point of view understanding the steady-state operation of the cell is key to its development as a useful photovoltaic device. In addition, we have demonstrated how the efficiency-determining parameters of the solar cell can be obtained from experimental data.

We were able to show that recombination in the bulk of the perovskite material combined with slow ionic motion is sufficient to reproduce the most common features of the impedance response for a wide variety of perovskite solar cell devices, including flat high frequency capacitance, low frequency negative capacitance features and inductive loops.

We found that the electrical properties, that determine device efficiency under operational conditions (i.e. at steady state), can be deduced solely from the high frequency impedance response. This is the case for the recombination resistance, the so-called "ideality factor" and the charge collection efficiency of the device. However, we also showed that these properties depend on the steady-state distribution of mobile ionic charge within the perovskite layer. It follows that, for a fair comparison to be made between different devices, the electrical properties should be deduced from high frequency impedance measurements performed at the values of the applied DC voltage for which the cells are expected to operate (i.e. close to the maximum power point in the IV curve). In contrast, the low frequency impedance response is primarily determined by the charging, and discharging, of the interfacial Debye layers in response to ion vacancy motion within the perovskite. The low frequency impedance response can thus be used as a tool to infer ionic mobility and the capacitances of the Debye layers.

In summary, by conducting a side-by-side comparison between theory and experiment, we have provided a sound basis with which to extract information about recombination and charge collection efficiency from actual impedance measurements in perovskite solar cells, regardless of the ionic characteristics of the perovskite material.

**Acknowledgements**

This work was funded by the United Kingdom's Engineering and Physical Sciences Research Council (EPSRC) through the Centre for Doctoral Training in New and Sustainable Photovoltaics (EP/L01551X). It also received funding from the European Union's Horizon 2020 research and innovation programme under the EoCoE II project (824158). JAA, AR, LCB thank Ministerio de Ciencia e Innovación of Spain, Agencia Estatal de Investigación (AEI) and EU (FEDER) under grants MAT2016-79866-R and PCI2019-111839-2 (SCALEUP) and Junta de Andalucía for support under grant SOLARFORCE (UPO-1259175) ). AR thanks the Spanish Ministry of Education, Culture and Sports via a PhD grant (FPU2017-03684). NEC was supported by an EPSRC

Doctoral Prize (ref. EP/R513325/1) LJB is supported by an EPSRC funded studentship from the CDT in New and Sustainable Photovoltaics, reference EP/L01551X/1

# Deducing the key physical properties of a perovskite solar cell from its impedance response: insights from drift-diffusion modelling
# – Supporting Information

Antonio Riquelme[a,#], Laurence J. Bennett[b,#], Nicola E. Courtier[b], Matthew J. Wolf[c], Lidia Contreras-Bernal[a], Alison Walker[c,*], Giles Richardson[b,*], Juan A. Anta[a,*]

[a] Área de Química Física, Universidad Pablo de Olavide, E-41013, Seville, Spain.

[b] Mathematical Sciences, University of Southampton, SO17 1BJ, UK.

[c] Department of Physics, University of Bath, BA2 7AY, UK

#Both authors contributed equally to this work

**Experimental details**

Perovskite solar devices with regular configuration (TiO$_2$/MAPbI$_3$/Spiro-OMeTAD)[1,2] were fabricated on FTO-coated glass (Pilkington–TEC15) patterned by laser etching. Before the deposition, the substrates were cleaned using Hellmanex® solution and rinsed with deionized water and ethanol. Thereupon, they were ultrasonicated in 2-propanol and dried by using compressed air. The TiO$_2$ blocking layer was deposited onto the substrates by spray pyrolysis at 450 °C, using a titanium diisopropoxide bis(acetylacetonate) solution (75% in 2-propanol, Sigma Aldrich) diluted in ethanol (1:14, v/v), with oxygen as carrier gas. The TiO$_2$ compact layer was then kept at 450 °C for 30 min for the formation of anatase phase. Once the samples achieve room temperature, a TiO$_2$ mesoporous layer was deposited by spin coating at 2000 rpm during 10 s using a commercial TiO$_2$ paste (Dyesol, 18NR-T) diluted in ethanol (1:5, weight ratio). After drying at 100 °C for 10 min, the TiO2 mesoporous layer was heated at 500 °C for 30 min and later cooled to room temperature. An additional doping treatment using Li+ ions (10.04 mg LiTFSI in 1 ml acetonitrile, 35mM) was used for the TiO2 mesoporous layer prior to CsPbBr3 deposition.

For MAPbI3 based devices, a pure methylammonium lead iodide solution were prepared to be deposited by spin coating using a methodology previously reported:[2] The perovskite precursor solution was adjusted to the relative humidity of the environment (42% R.H.) by the Pb/DMSO ratio. The perovskite precursor solution (50 µL) was spin-coated in a one-step setup at 4000 rpm for 50 s. During this step, DMF is selectively washed with non-polar diethyl ether just before the white solid begins to crystallize in the substrate. Spiro-OMeTAD was deposited as hole selective material by dissolving 72.3 mg in 1 mL of chlorobenzene as well as 17.5 µL of a lithium bis (trifluoromethylsulphonyl)imide (LiTFSI) stock solution (520 mg of LiTFSI in 1mL of acetonitrile), and 28.8 µL of 4-tert-butylpyridine (TBP). The Spiro-OMeTAD was spin coated at 4000 rpm for 30 s. The solution was filtered with a 0.2 µm PTFE filter prior to their deposition. Finally, 60 nm of gold was deposited as a metallic contact by thermal evaporation under a vacuum level between $1\cdot 10^{-6}$ and $1\cdot 10^{-5}$ torr. All the deposition processes were carried out outside a glovebox under environmental conditions.

**Numerical Model**

The simulated device consists of an electron transport layer, perovskite absorber layer and hole transport layer: ETL/MAPbI3/HTL.[3] Within the perovskite layer ($0 < x < b$), the electron and hole densities, $n$ and $p$ respectively, are governed in time, $t$, and one spatial dimension, $x$, via the continuity equations

$$\frac{\partial n}{\partial t} - \frac{1}{q}\frac{\partial j_n}{\partial x} = G - R, \qquad j_n = qD_n\left(\frac{\partial n}{\partial x} - \frac{n}{k_BT}\frac{\partial \phi}{\partial x}\right), \tag{1}$$

$$\frac{\partial p}{\partial t} + \frac{1}{q}\frac{\partial j_p}{\partial x} = G - R, \qquad j_p = -qD_p\left(\frac{\partial p}{\partial x} + \frac{p}{k_BT}\frac{\partial \phi}{\partial x}\right), \tag{2}$$

where $q$ is the elementary charge, $G$ and $R$ the generation and recombination rates respectively, $D_{n,p}$ the respective diffusion coefficients, $k_B$ the Boltzmann constant and $T$ is the temperature. The migration of the mobile anion vacancy density, $P$, within the perovskite layer is modelled by

$$\frac{\partial P}{\partial t} + \frac{\partial F_P}{\partial x} = 0, \qquad F_P = -D_P\left(\frac{\partial P}{\partial x} + \frac{P}{k_BT}\frac{\partial \phi}{\partial x}\right), \tag{3}$$

where $F_P$ is the ionic flux and $D_P$ is the anion vacancy diffusion coefficient. Equations (1-3) couple to Poisson's equation for the electric potential

$$\frac{\partial^2 \phi}{\partial x^2} = \frac{q}{\epsilon_p}(N_0 - P + n - p). \tag{4}$$

Here, $\epsilon_p$ is the permittivity of the perovskite and $N_0$ is the uniform cation vacancy density which is equal to the mean anion vacancy density to ensure charge neutrality. Electrons and holes are generated within the perovskite via the Beer-Lambert profile for a single wavelength of light

$$G(x) = F_{ph}\alpha e^{-\alpha x}, \tag{5}$$

where $F_{ph}$ is the flux of photons incident on the device with energy above the bandgap and $\alpha$ is the absorption coefficient of the perovskite. Recombination within the bulk of the perovskite is calculated using a combination of bimolecular (direct relaxation across the bandgap) and trap-assisted Shockley-Read-Hall (SRH) schemes, given by

$$R(n,p) = \beta(np - n_i^2) + \frac{(np - n_i^2)}{\tau_n p + \tau_p n + k_3}, \tag{6}$$

where $\tau_n$ and $\tau_p$ are the SRH pseudo-lifetimes for electrons and holes respectively and $k_3$ is a constant from the deep trap approximation used by Courtier et al.[3] The intrinsic carrier density $n_i$ is defined by

$$n_i = g_c g_v \exp\left(-\frac{E_G}{2k_B T}\right). \tag{7}$$

Here, $g_c$ and $g_v$ are the density of states in the conduction and valence bands respectively and $E_G$ is the bandgap of the perovskite. Interfacial recombination is determined at the ETL/perovskite and the perovskite/HTL interfaces via

$$R_l(n,p) = \frac{n|_{x=0^-}p|_{x=0^+} - n_i^2}{p|_{x=0^+}/v_{n_E} + n|_{x=0^-}/v_{p_E} + k_1}, \tag{8}$$

where $v_{n_E}$ and $v_{p_E}$ are the electron and hole recombination velocities at the ETL/perovskite interface and $k_1$ is a constant from the deep trap approximation at the ETL/perovskite interface. The electron concentration in the ETL at the interface and the hole concentration in the perovskite at the interface are denoted by $n|_{x=0^-}$ and $p|_{x=0^+}$, respectively. Similarly, at the perovskite/HTL interface

$$R_r(n,p) = \frac{n|_{x=b^-}p|_{x=b^+} - n_i^2}{p|_{x=b^+}/v_{n_H} + n|_{x=b^-}/v_{p_H} + k_2}, \tag{9}$$

where $v_{n_H}$ and $v_{p_H}$ are the electron and hole recombination velocities at the perovskite/HTL interface and $k_2$ is a constant from the deep trap approximation at the perovskite/HTL interface. The electron concentration in the perovskite at the interface and the hole concentration in the HTL at the interface are denoted by $n|_{x=b^-}$ and $p|_{x=b^+}$ respectively.

The anion vacancies are limited to the perovskite layer as well as generation and bulk recombination. Hence the conservation equations for electrons in the ETL ($-b_E < x < 0$) are given by

$$\frac{\partial n}{\partial t} - \frac{1}{q}\frac{\partial j_n}{\partial x} = 0, \qquad j_n = qD_E\left(\frac{\partial n}{\partial x} - \frac{n}{k_B T}\frac{\partial \phi}{\partial x}\right), \qquad (9)$$

with

$$\frac{\partial^2 \phi}{\partial x^2} = \frac{q}{\epsilon_E}(n - d_E), \qquad (10)$$

where $D_E$ is the electronic diffusion coefficient in the ETL and $\epsilon_E$ and $d_E$ are the permittivity and effective doping density of the ETL respectively. Similarly, the conservation equations for holes in the HTL ($b < x < b + b_H$) are given by

$$\frac{\partial p}{\partial t} + \frac{1}{q}\frac{\partial j_p}{\partial x} = 0, \qquad j_p = -qD_H\left(\frac{\partial p}{\partial x} + \frac{p}{k_B T}\frac{\partial \phi}{\partial x}\right), \qquad (11)$$

with

$$\frac{\partial^2 \phi}{\partial x^2} = \frac{q}{\epsilon_H}(d_H - p). \qquad (12)$$

Here, $D_H$ denotes the hole diffusion coefficient in the HTL and $\epsilon_H$ and $d_H$ are the HTL permittivity and effective doping density respectively. A number of boundary conditions are applied to equations (1-12) to simulate the physics of PSC operation. To model metal contacts at the ETL and HTL edges, we apply

$$\phi|_{x=-b_E} = 0, \quad n|_{x=-b_E} = d_E, \quad \phi|_{x=b+b_H} = V_{ap} - V_{bi}, \quad p|_{x=b+b_H} = d_H. \qquad (13)$$

At the ETL/ MAPbI₃ interface

$$\phi|_{x=0^-} = \phi|_{x=0^+}, \qquad \epsilon_E \frac{\partial \phi}{\partial x}\bigg|_{x=0^-} = \epsilon_p \frac{\partial \phi}{\partial x}\bigg|_{x=0^+}, \qquad k_E n|_{x=0^-} = n|_{x=0^+},$$

$$j_p|_{x=0} = -qR_{I_l}, \qquad F_p|_{x=0} = 0. \qquad (14)$$

At the MAPbI₃/HTL interface

$$\phi|_{x=b^-} = \phi|_{x=b^+}, \qquad \epsilon_p \frac{\partial \phi}{\partial x}\bigg|_{x=b^-} = \epsilon_H \frac{\partial \phi}{\partial x}\bigg|_{x=b^+}, \qquad p|_{x=b^-} = k_H p|_{x=b^+},$$

$$j_n|_{x=b} = -qR_{I_r}, \qquad F_p|_{x=0} = 0. \qquad (15)$$

Here, $k_E$ and $k_H$ specify the ratio of the carrier densities at either side of the perovskite and transport layer interface. They are defined by

$$k_E = \frac{g_c}{g_{c,E}}\exp\left(-\frac{E_c-E_{c,E}}{qk_bT}\right), \qquad k_H = \frac{g_v}{g_{v,H}}\exp\left(-\frac{E_v-E_{v,H}}{qk_bT}\right), \qquad (16)$$

where $g_c$ and $g_v$ are the density of states of the perovskite conduction and valance bands respectively, $g_{c,E}$ and $g_{v,H}$ are the density of states of the ETL conduction band and HTL valence band respectively, $E_c$ and $E_v$ are the energies of the conduction and valence bands of the perovskite and $E_{c,E}$ and $E_{v,H}$ are the energies of the conduction and valence bands of the ETL and HTL respectively.

For this work, it is assumed that the density of states of the ETL and HTL are equal to the effective doping densities in those layers, i.e. $g_{c,E} = d_E$ and $g_{v,H} = d_H$. As a result, the band edges $E_{c,E}$ and $E_{v,H}$ are equal to the Fermi levels in the ETL and HTL, denoted by $E_{fE}$ and $E_{fH}$, respectively. Hence, the built-in voltage is given by

$$V_{bi} = \frac{1}{q}(E_{fE} - E_{fH}) = \frac{1}{q}(E_{c,E} - E_{v,H}). \qquad (17)$$

An identical model was previously used by Courtier et al.[3]

Equations (1-12) with boundary conditions (13-15) are solved using the open-source numerical solver *IonMonger*.[4]

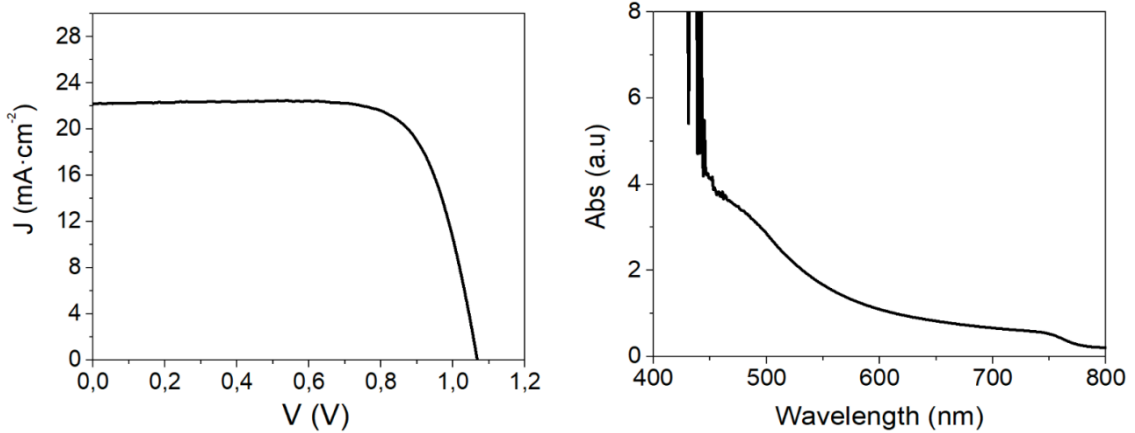

**Figure S1. (a)** Experimental JV curve at 1 Sun (1000 W/m²) AM1.5G illumination for perovskite solar devices with regular configuration (TiO$_2$/MAPbI$_3$/Spiro-OMeTAD), **(b)** absorption spectrum of the same device measured at transmittance mode.

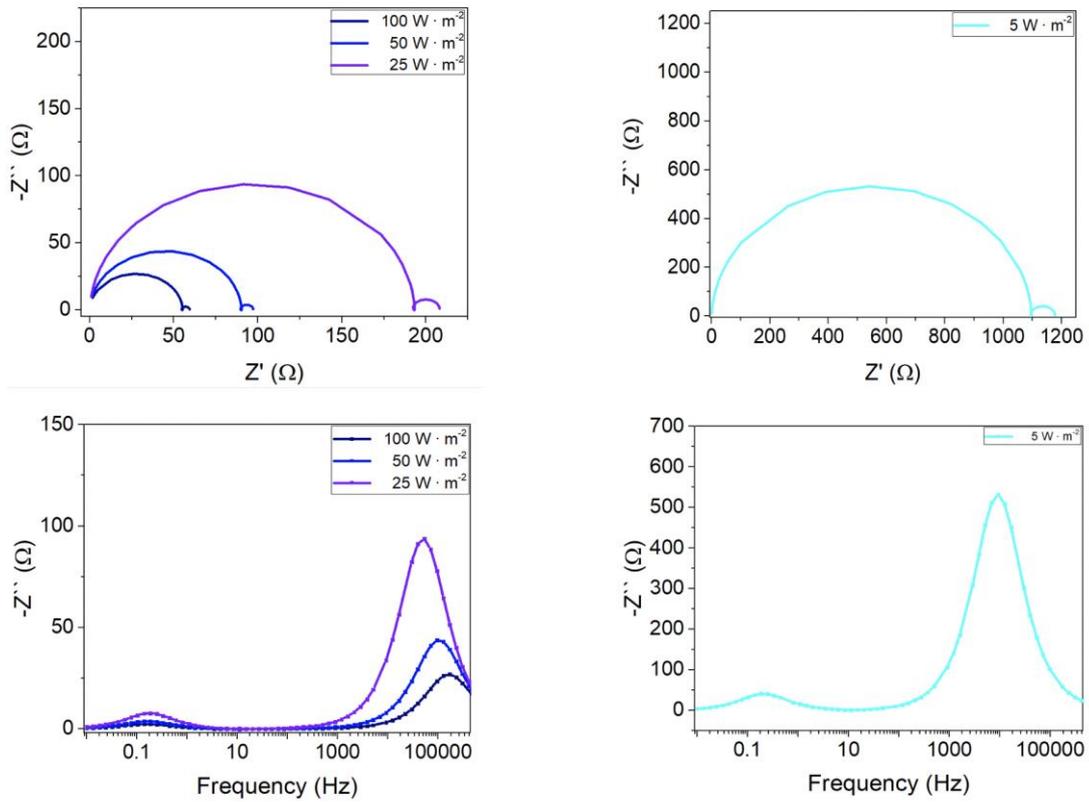

**Figure S2.** Simulated impedance spectra at open-circuit and variable illumination intensity for perovskite solar devices with regular configuration (TiO2/MAPbI3/Spiro-OMeTAD) using the parameters indicated in Table 1.

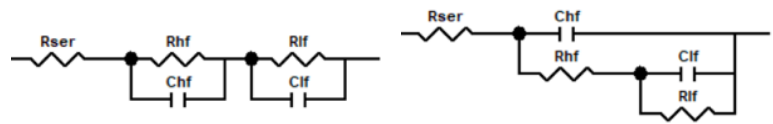

**Figure S3.** Equivalent circuits used to fit both the experimental and the simulated impedance data.

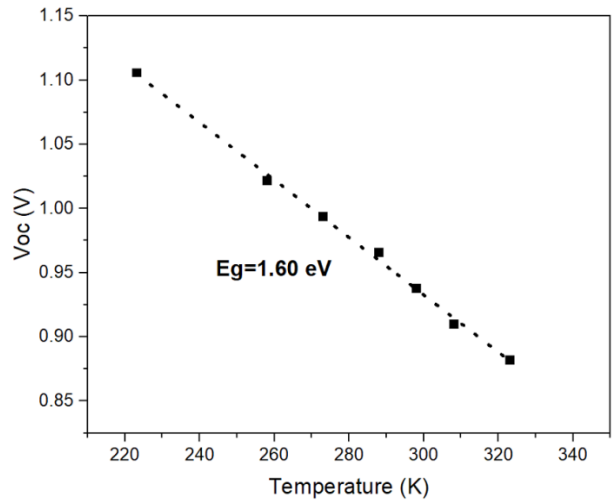

**Figure S4.** DD numerical prediction of the temperature dependence of the open-circuit photovoltage for the standard parameters of Table 1.

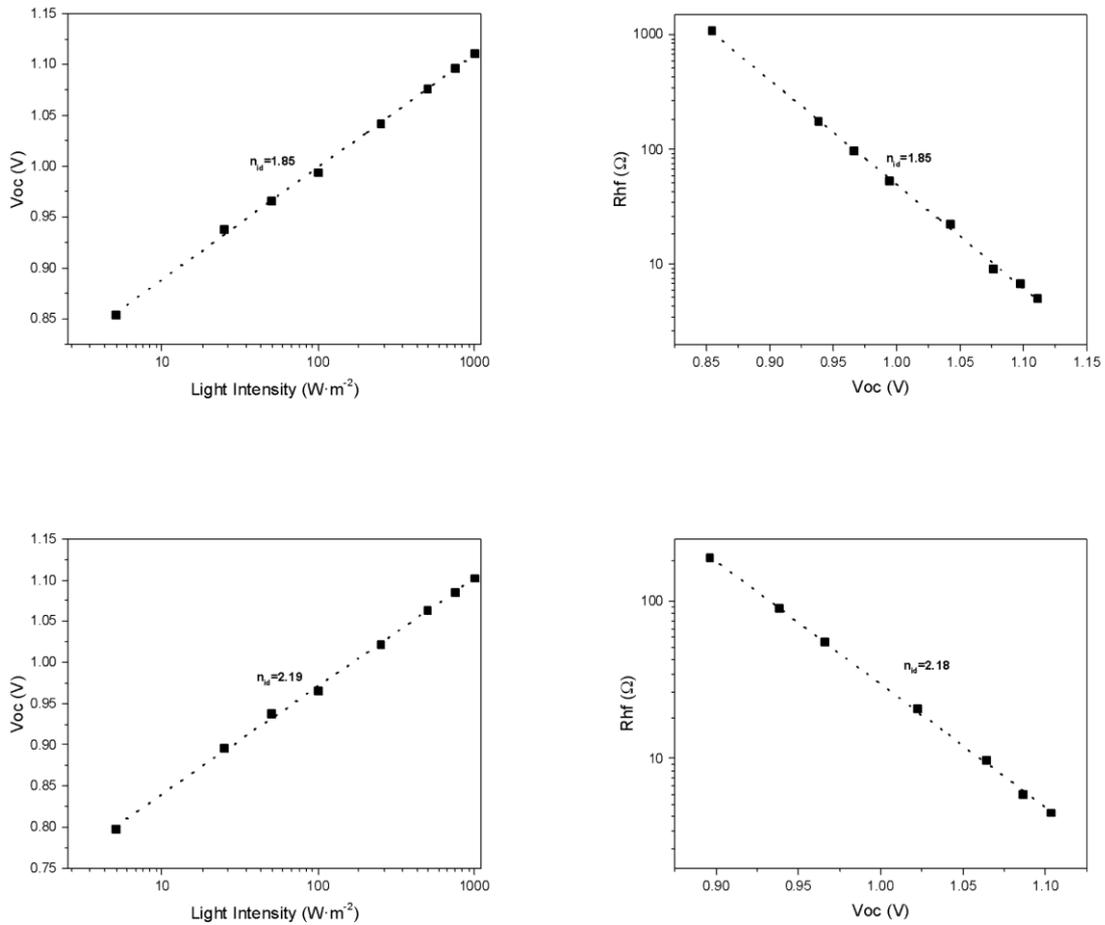

**Figure S5.** DD numerical prediction of the ideality factor from high frequency resistance and open circuit voltage for two different values of the density of vacancies. **Top**: $10^{25}$ m$^{-3}$, **bottom**: 1.8 $10^{25}$ m$^{-3}$

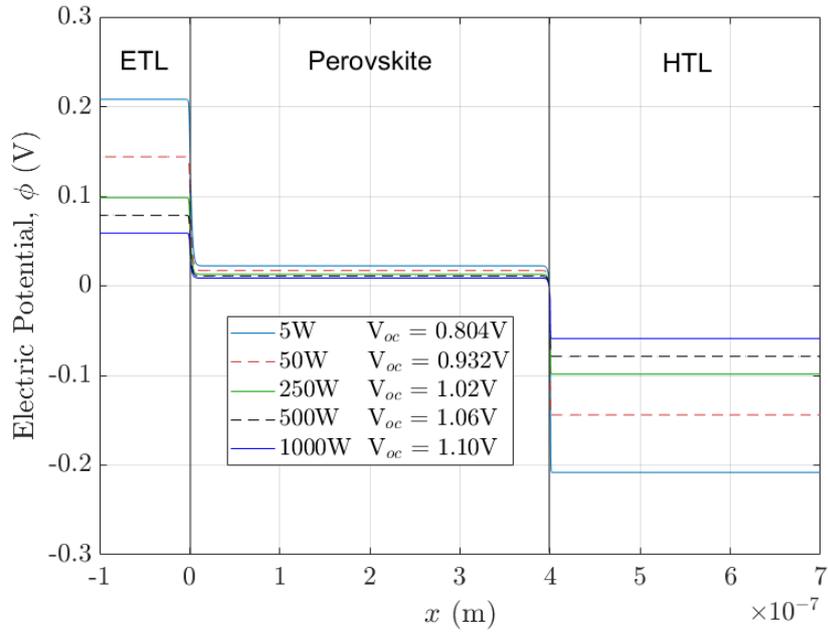

**Figure S6.** Steady state electric potentials across the cell at different open-circuit voltages with the standard parameter set (Table 1)

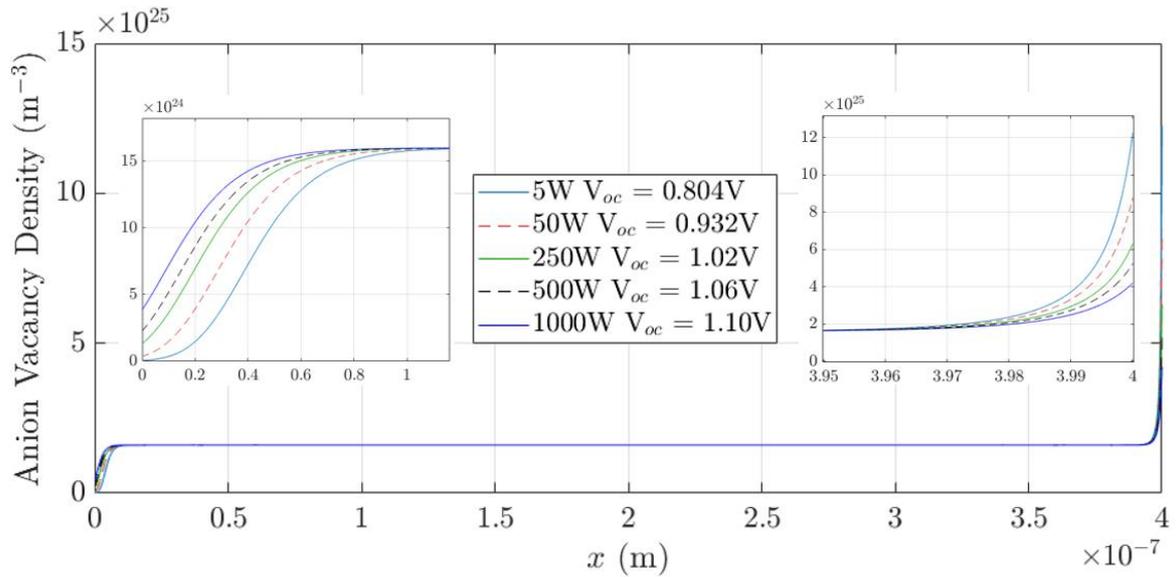

**Figure S7.** Steady state anion vacancy densities at different open-circuit voltages with the standard parameter set (Table 1). Insets show the left and right perovskite Debye layers.

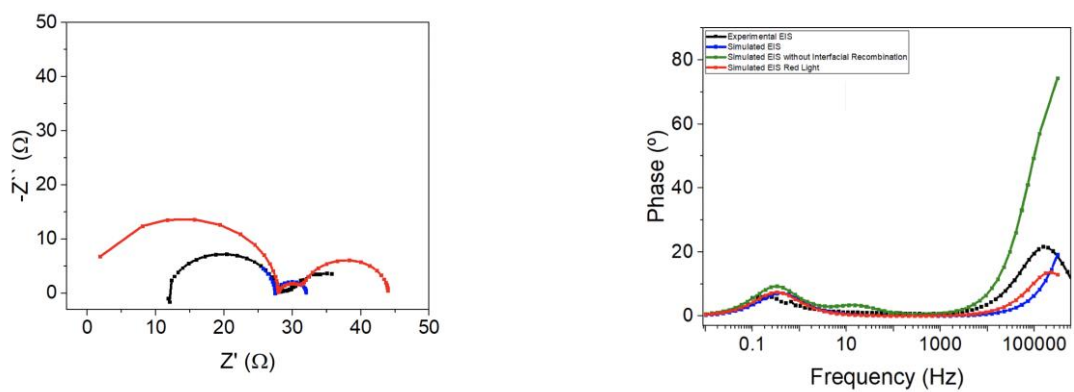

**Figure S8.** Simulated and experimental impedance spectra at open-circuit for perovskite solar devices with regular configuration (TiO2/MAPbI3/Spiro-OMeTAD) using the parameters indicated in Table 2.

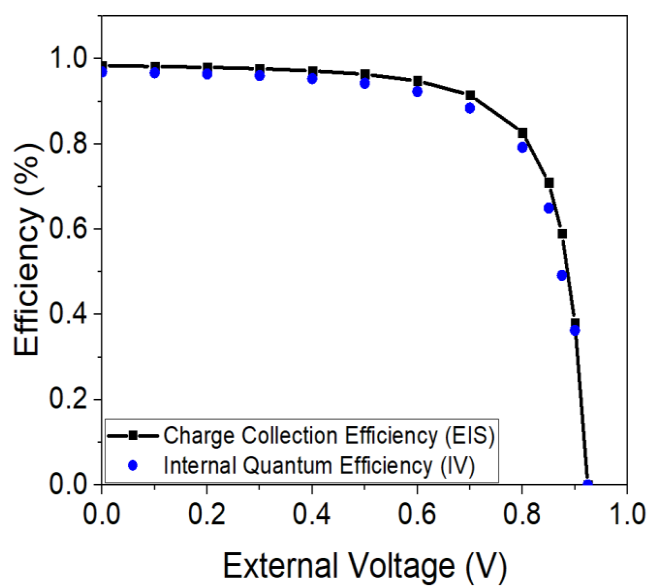

**Figure S9.** Simulated charge collection efficiency as predicted from Eq. (13) and the impedance parameters of Figure 6b when we assume the identity Rrec = RHF (black line). The stationary internal quantum efficiency derived from the simulated JV curve (current density / q * absorbed photon flux) is included in the graph (blue circles). Simulation parameters are those in Table 1.